\documentclass[a4paper,11pt]{article}
\pdfoutput=1 
\usepackage{jcappub} 
\usepackage[T1]{fontenc} 
\usepackage{physics}
\usepackage{wrapfig}
\usepackage{xcolor}
\usepackage{float}
\restylefloat{table}
\usepackage{changepage}
\usepackage{booktabs}
\usepackage{subcaption}
\usepackage{graphicx}
\usepackage{csquotes}
\usepackage[normalem]{ulem}

\newcommand{\CNB}{C\texorpdfstring{$\nu$}{nu}B }

\title{Numerical treatment of annual modulation of relic neutrinos}

\author[a,1]{Fabian Zimmer,\note{Corresponding author.}}
\author[a,b]{and Shin'ichiro Ando}

\affiliation[a]{GRAPPA Institute, University of Amsterdam, Science Park 904, 1098 XH Amsterdam, The Netherlands}
\affiliation[b]{Kavli Institute for the Physics and Mathematics of the Universe, University of Tokyo, Chiba 277-8583, Japan}

\emailAdd{f.zimmer@uva.nl}
\emailAdd{s.ando@uva.nl}

\abstract{
We present a numerical treatment of the annual modulation of relic neutrinos due to the Sun's gravitational influence. Extending our previously developed N-1-body simulation framework from Milky Way scales to solar system dynamics, we model how cosmic neutrino background densities might vary throughout Earth's orbital cycle. We validate our numerical approach against analytical expectations from previous studies that assumed idealized relic neutrino populations. Our results suggest that the prior gravitational history of neutrinos traversing asymmetric dark matter distributions can affect annual modulation patterns. While our simulations reproduce modulation amplitudes similar to analytical predictions for heavier neutrinos, we find that the amplitude can vary considerably depending on the specific morphology of dark matter halos. These findings highlight the importance of incorporating realistic structure formation effects when predicting potential observational relic neutrino signatures.
}


\begin{document}
\maketitle
\flushbottom

\section{Introduction}
\label{sec:intro} 

The Cosmic Neutrino Background (C$\nu$B) represents one of the most fundamental yet elusive components of our Universe. Despite being predicted to be the second most abundant particle species after photons, the \CNB still evades direct laboratory detection and its properties have also not been indirectly detected even with the newest generation of high-precision, large-scale cosmological surveys \cite{DESI:2024mwx}.

The gravitational influence on relic neutrinos manifests differently across cosmic scales, creating a rich phenomenology that has only recently begun to be explored in detail. On the largest cosmological scales, neutrinos are gravitationally attracted to dark matter (DM) and baryonic structures, with each component subsequently enhancing the growth of the other \cite{Yu:2016yfe}. As we transition to galactic scales, non-linear gravitational effects become increasingly important and require sophisticated numerical modeling to properly capture the clustering dynamics of relic neutrinos \cite{Ringwald:2004np,deSalas:2017wtt,Zhang:2017ljh,Mertsch:2019qjv,Elbers:2023mdr,Zimmer:2023jbb,Zimmer:2024max}. On even smaller scales within our solar system, analytical studies have demonstrated that the Sun's gravitational field can focus the \CNB flux incident on Earth, creating a time-dependent modulation signal throughout Earth's orbital period \cite{Safdi:2014rza}. This annual modulation phenomenon could provide an observable signature that might aid in the eventual detection of the \CNB. The effect arises because Earth's varying position relative to the Sun throughout its orbit alters the gravitational focusing conditions, leading to seasonal variations in the local neutrino flux. Understanding and accurately predicting this modulation may become important as detection experiments continue to develop.

Indeed, the experimental landscape for \CNB detection is growing, with various proposed approaches being in different theoretical or developmental stages. The PTOLEMY (Princeton Tritium Observatory for Light, Early-Universe, Massive-Neutrino Yield) concept \cite{Cocco:2007za, Long:2014zva,Betts:2013uya,PTOLEMY:2018jst,PTOLEMY:2019hkd} represents one of the most studied approaches, proposing to utilize tritium beta decay spectra to search for signals of relic neutrino capture. Such tritium-based approaches might be enhanced by using Cyclotron Radiation Emission Spectroscopy (CRES) technology \cite{Iwasaki:2024voi} pioneered by the Project8 collaboration \cite{Project8:2014ivu,Project8:2017nal,Project8:2022hun,Project8:2023jkj}, although overall technical challenges remain \cite{PTOLEMY:2022ldz}. 

It is interesting to note the increasingly creative detection strategies devised by researchers over the years, including atomic de-excitation spectroscopy that could be sensitive to Pauli blocking by ambient relic neutrinos \cite{Takahashi:2007ec,Yoshimura:2014hfa}, schemes using accelerated ions to leverage enhanced neutrino interaction cross-sections \cite{Bauer:2021uyj}, and detectors that might exploit the large de Broglie wavelength of relic neutrinos \cite{Arvanitaki:2022oby,Arvanitaki:2023fij}, among other speculative ideas \cite{Stodolsky:1974aq,Dev:2021tlo,2023arXiv231003315A,Chauhan:2024deu,Das:2024thc,Hernandez-Molinero:2025vor}.

Tritium-based experiments like PTOLEMY may also leverage annual modulation to aid \CNB detection despite lower experimental energy resolution \cite{Akhmedov:2019oxm}, or to distinguish between thermal and non-thermal relic neutrino populations \cite{Huang:2016qmh}. The latter may arise from alternative cosmological scenarios \cite{Chen:2015dka}, where the resulting relic neutrinos typically exhibit lower average velocities \cite{Trautner:2016ias}, potentially enhancing gravitational focusing effects.

Another relevant avenue of research involves using the \CNB as a scattering medium and exploring the visibility of associated observables in future neutrino experiments like Hyper-Kamiokande or IceCube-Gen2 (see e.g. \cite{Brdar:2022kpu,Wang:2025qap,Balantekin:2023jlg,Esteban:2021tub}). Such studies rely on assumed \CNB number densities and phase-space distributions, usually neglecting gravitational clustering effects, and would therefore benefit from dedicated numerical investigations of (local) \CNB variations.

While the analytical framework developed in \cite{Safdi:2014rza} provided crucial insights into solar gravitational focusing, it necessarily relied on simplified assumptions about the \CNB phase-space distribution and neglected the prior gravitational processing of neutrinos by Milky-Way (MW)-scale structure. Our previous work has demonstrated that galactic-scale gravitational clustering can modify the local neutrino distribution \cite{Zimmer:2023jbb,Zimmer:2024max}, sometimes creating unexpected underdensities even where analytical treatments predict only enhancements. This raises important questions about how the complex interplay between galactic and solar gravitational effects influences the annual modulation signal.

In this work, we address these questions by extending our established N-1-body simulation framework to provide the first comprehensive numerical treatment of \CNB annual modulation. Our approach uniquely combines the gravitational clustering effects computed over cosmological timescales with accurate modeling of solar system dynamics throughout Earth's orbital cycle. This dual-scale treatment allows us to investigate how the prior gravitational history of relic neutrinos — including their interactions with asymmetric DM halo structures — affects the annual modulation patterns that might be observable at Earth. We can also treat the \CNB as a unified population, rather than assuming idealized fully-bound or fully-unbound populations as in \cite{Safdi:2014rza}. We validate our numerical framework against established analytical predictions before exploring the phenomenology that emerges when incorporating the full complexity of structure formation effects. Such a dual treatment was done, as far as we are aware, only with respect to direct DM detection rates: analytically in \cite{Griest:1987vc,Lee:2013wza} and numerically in \cite{Alenazi:2006wu}. 

The paper is organized as follows: Section~\ref{sec:methods} details our numerical methodology, reviewing both the analytical framework and our simulation approach for modeling Earth-Sun orbital dynamics. Section~\ref{sec:results} presents our results, including validation against analytical predictions and exploration of how different DM halo morphologies affect modulation amplitudes. We conclude and discuss our results in section~\ref{sec:discussion_and_conclusions}.

\section{Methods}
\label{sec:methods}

This section outlines our approach to studying \CNB annual modulation. We first review the analytical method developed by \cite{Safdi:2014rza}, followed by a short description of our numerical simulation framework. Finally, we detail our implementation of Earth-Sun orbital dynamics based on \cite{Lee:2013xxa,McCabe:2013kea}, which enables us to model solar gravitational focusing throughout Earth's orbit. Effectively, we combine results from our core simulations of \cite{Zimmer:2023jbb} with a simple, dedicated simulation setup for solar focussing of the \CNB, which allows us to examine potential nuances in \CNB modulation that may emerge when considering both galactic and solar gravitational effects.

\subsection{Analytical method}
\label{sec:ana_methods}

The analytical method uses a few simplifications. First, the gravitational potential of the MW is not taken into account, only the gravity of the Sun enters the framework. Second, a certain velocity distribution for the \CNB has to be assumed. As we do not know the exact phase-space distribution of the \CNB today, it is assumed that all neutrinos are in a fully gravitationally unbound or bound state.

Even though the \CNB is largely non-relativistic today, the unbound state can be described by the isotropic Fermi-Dirac distribution
\begin{equation}
    f_{\rm FD}(p_\nu) = \frac{1}{1+e^{p_\nu / T_\nu}} ,
\end{equation}
which holds even after neutrino decoupling due to particle number conservation. For the bound state, it is assumed that the \CNB had time to fully virialise, i.e. is fully gravitationally bound to the MW. Their velocity distribution is then isotropic in the Galactic frame, and treated to follow the velocity distribution of the DM particles comprising the MW halo, which is typically modeled by the standard halo model (SHM)
\begin{equation}
    f_{\rm SHM}\left(\mathbf{v}_\nu\right) = \begin{cases}\frac{1}{N_{\mathrm{esc}}}\left(\frac{1}{\pi v_0^2}\right)^{3 / 2} e^{-\mathbf{v}_\nu^2 / v_0^2} & \left|\mathbf{v}_\nu\right|<v_{\mathrm{esc}} \\ 0 & \left|\mathbf{v}_\nu\right| \geq v_{\mathrm{esc}}\end{cases} ,
\end{equation}
where the cutoff represents the escape velocity $v_{\mathrm{esc}} \approx 550 \mathrm{~km} / \mathrm{s}$ \cite{Smith:2006ym} and $v_0$ the velocity dispersion of this distribution. The normalization factor includes the error function (abbreviated as erf) and is given by (see e.g. \cite{Lee:2013xxa})
\begin{equation}
    N_{\mathrm{esc}}=\operatorname{erf}(z)-\frac{2}{\sqrt{\pi}} z e^{-z^2}, \quad z \equiv \frac{v_{\mathrm{esc}}}{v_0} .
\end{equation}
Both distributions are not yet in Earth's frame, the frame in which we would like to know how the \CNB density is modulated over the year. For this end, Liouville's theorem is used to equate the phase-space densities at Earth's location with one asymptotically far away from the Earth-Sun system. For clarity, we will describe each scenario (fully unbound and fully bound) separately.

\subsubsection{Unbound case}
For the unbound case we can write Liouville's theorem as
\begin{equation}
    f_{\oplus}\left(\mathbf{v}_\nu , t \right) = f_{\rm FD} \left(\mathbf{v}_{\mathrm{C\nu B}}+\mathbf{v}_{\infty}\left[\mathbf{v}_\nu+\mathbf{V}_{\oplus}(t)\right]\right)
\end{equation}
to connect the phase-space at Earth's location, $f_{\oplus}$, of neutrino velocities $\mathbf{v}_\nu$ with the unbound distribution, $f_{\rm FD}$.
The velocity $\mathbf{v}_{\mathrm{C\nu B}}$ appears due to the motion of the Earth-Sun system relative to the C$\nu$B. 
We adopt the typical assumption that the \CNB and CMB rest frames coincide. Therefore, the Sun's velocity relative to the \CNB is the same as relative to the CMB, and we can use the associated velocity of the measured CMB dipole of $\mathbf{v}_{\mathrm{C\nu B}} \approx (-0.0695, -0.662, 0.747) \times 369 \mathrm{~km} / \mathrm{s}$ \cite{Planck:2018vyg} in our simulations.
The other velocity term given by \cite{Alenazi:2006wu}
\begin{equation}
    \mathbf{v}_{\infty}\left[\mathbf{v}_{\mathbf{s}}\right]=\frac{v_{\infty}^2 \mathbf{v}_{\mathbf{s}}+v_{\infty}\left(G M_{\odot} / r_s\right) \hat{\mathbf{r}}_{\mathbf{s}}-v_{\infty} \mathbf{v}_{\mathbf{s}}\left(\mathbf{v}_{\mathbf{s}} \cdot \hat{\mathbf{r}}_{\mathbf{s}}\right)}{v_{\infty}^2+\left(G M_{\odot} / r_s\right)-v_{\infty}\left(\mathbf{v}_{\mathbf{s}} \cdot \hat{\mathbf{r}}_{\mathbf{s}}\right)}
\end{equation}
and represents the necessary initial velocity of a particle in the Solar frame to have a velocity $\mathbf{v}_{\mathbf{s}}$ at Earth's location, where we have $v_{\infty}^2=v_s^2-2 G M_{\odot} / r_s$ due to energy conservation, and $\mathbf{r}_{\mathbf{s}}$ represents Earth's position vector relative to the Sun. Lastly, $\mathbf{V}_{\oplus}(t)$ is Earth's velocity relative to the Sun, which varies slightly throughout the year \cite{Lee:2013xxa,McCabe:2013kea}. In section \ref{sec:num_methods} we discuss the last two quantities in more detail. Using the non-relativistic approximation $p \approx m_\nu v_\nu$ and integrating over velocity gives the time-dependent number density at Earth
\begin{equation}
    n_{\nu,\oplus}(t) = \int m_\nu^3  f_{\oplus}\left( v_\nu , t \right) \frac{d^3 v_\nu}{(2 \pi)^3} .
\label{eq:n_nu_unbound}
\end{equation}

\subsubsection{Bound case}
The cutoff condition in the velocity distribution of the SHM models the fact that less of a lighter neutrino population (which has higher average velocities) is bound to the MW as integrating over $f_{\rm SHM}$ with higher velocities will result in a value $\ll 1$. Liouville's theorem for this case is
\begin{equation}
    f_{\oplus}\left(\mathbf{v}_\nu\right) = f_{\rm SHM} \left(\mathbf{v}_{\mathrm{MW}}+\mathbf{v}_{\infty}\left[\mathbf{v}_\nu+\mathbf{V}_{\oplus}(t)\right]\right) .
\end{equation}
This time, the germane relative velocity is that of the Sun in the MW Galactic frame of $\mathbf{v}_{\mathrm{MW}} \approx (0.0473, 0.9984, 0.0301) \times 232 \mathrm{~km} / \mathrm{s}$ \cite{Planck:2018vyg}. The number density on Earth is then obtained by integrating as
\begin{equation}
    n_{\nu,\oplus}(t) = n_{\nu,0} \int f_{\oplus}\left(v_\nu\right) \frac{d^3 v_\nu}{(2 \pi)^3},
\label{eq:n_nu_bound}
\end{equation}
where $n_{\nu,0} \approx 56 {\rm ~cm}^{-3}$ is the number density per neutrino family, per helicity as predicted by $\Lambda$CDM cosmology. This additional factor is needed, as the distribution in this case is bounded between $0$ and $1$.

To clarify the physical mechanism, both distributions $f_{\rm FD}$ and $f_{\rm SHM}$ are bounded, but the presence of $\mathbf{v}_{\infty}$ can map neutrino velocities to regions of the distribution where the phase-space density $f(v)$ is either higher or lower than would be obtained without solar gravitational effects. Consequently, when integrating to compute the total number density, the result can exceed $n_{\nu,0}$ if gravitational focusing predominantly samples high-valued regions of the distribution, or fall below $n_{\nu,0}$ if it samples low-valued regions.

To obtain the annual modulation, we use the definition
\begin{equation}
    \mathcal{M}(t) \equiv \frac{n_\nu(t) - n_{\nu,{\rm min}}}{n_\nu(t) + n_{\nu,{\rm min}}} ,
\end{equation}
where $n_{\nu,{\rm min}}$ is the lowest number density of the year.

\subsection{Numerical method}
\label{sec:num_methods}

In previous works we have built a simulation framework to compute the \CNB densities at Earth's location today. We briefly describe the core aspects of the method here, while any omitted details can be found in \cite{Zimmer:2023jbb,Zimmer:2024max}.

\subsubsection{Core \CNB simulations}
Within our simulations, relic neutrinos simply follow Hamilton's equations of motion
\begin{equation} \label{eq:EOMs}
\mathbf{q}_j = a m_\nu \mathbf{x}_j', \quad \mathbf{q}_j'= -a m_\nu \frac{\partial \Phi(\mathbf{x}, t)}{\partial \mathbf{x}_j},
\end{equation}
with $\mathbf{q}_j$ and $\mathbf{x}_j$ being the comoving momentum and position of the $j$-th neutrino, respectively, $a$ the scale factor and $m_\nu$ the neutrino mass. Prime notation represents conformal time derivative. Effects from non-linear structure formation enter via the gravitational potentials $\Phi$ at position $\mathbf{x}$ and time $t$, which are formed by MW-like DM halos selected from the {\sc TangoSIDM} project \cite{Correa:2022dey,Correa:2024vgl}. Our available sample consists of 26 such halos, each represented by 25 snapshots spanning redshifts 0 to 4 (this last redshift here being labeled with $z_{\rm sim}$) to capture their evolution.
The simulation proceeds backward in time, with neutrinos initialized at Earth-equivalent galactic distances and aimed in different directions corresponding to the pixels of a healpy allsky map with a resolution of $\mathrm{Nside}=8$ (768 pixels).
Using Liouville's theorem we compute the present values of the phase-space as
\begin{equation} \label{eq:Liouville}
f(\mathbf{x}_\oplus, \mathbf{q}_j(0), 0) = f_{\rm sim} (\mathbf{x}_j(z_{\rm sim}), \mathbf{q}_j(z_{\rm sim}), z_{\rm sim}),
\end{equation}
where $\mathbf{x}_\oplus$ is the starting coordinate of all neutrinos and $f_{\rm sim}$ represents the initial phase-space distribution at $z_{\rm sim}$, which is assumed to be Fermi-Dirac like in \cite{Zimmer:2023jbb} or a slight deviation thereof due to evolving density anisotropies of the early Universe in \cite{Zimmer:2024max}. The local number density per degree of freedom is then computed as
\begin{equation} \label{eq:n_nu}
n_\nu = \int \frac{d^3 \mathbf{q}_0}{(2 \pi)^3} \, f(\mathbf{x}_\oplus, \mathbf{q}_0, 0),
\end{equation}
where the integral is evaluated as a sum over the momenta $\mathbf{q}_j(0)$ of all simulated neutrinos.

\subsubsection{Earth-Sun simulations}

\begin{figure}[t!]
    \centering
    \includegraphics[width=0.6\textwidth]{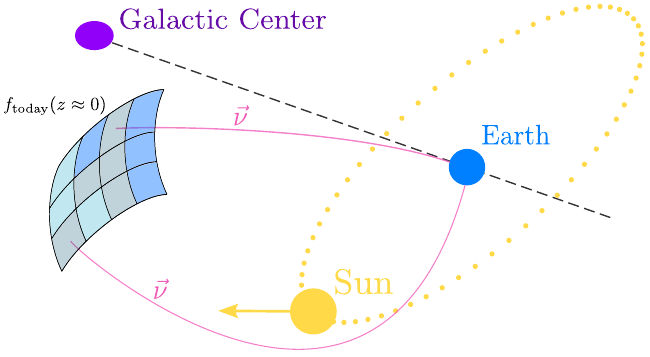}
    \caption{Illustration of our simulation setup. Neutrinos propagate back in time from Earth in a Milky Way-oriented frame. The Sun's positions throughout Earth's orbit are indicated by yellow dots. The colored grid represents phase-space values from our Galactic-scale simulations \cite{Zimmer:2023jbb}, which serve as the boundary conditions for the daily simulations. This approach allows us to incorporate both large-scale gravitational clustering effects and solar gravitational focusing within a unified framework.}
    \label{fig:mod_sketch}
\end{figure}

The aim of this work is to investigate, via dedicated numerical simulations, how the \CNB density gets modulated by the Sun's gravitational influence throughout one Earth orbit. For any omitted details in the following brief overview on the implementation of the Earth-Sun system see \cite{McCabe:2013kea,Lee:2013xxa}.

The Earth's orbit remains, to an excellent approximation, in its ecliptic plane. Therefore, Earth's relative position to the sun, formulated with the ecliptic unit vectors $\hat{\boldsymbol{\epsilon}}_{\mathbf{x}}$ and $\hat{\boldsymbol{\epsilon}}_{\mathbf{y}}$, is given as
\begin{equation}
\mathbf{r}_{\mathbf{s}} = r \cos \ell \hat{\boldsymbol{\epsilon}}_{\mathbf{x}} + r \sin \ell \hat{\boldsymbol{\epsilon}}_{\mathbf{y}} , \quad r = \frac{a\left(1-e^2\right)}{1 + e \cos \nu} ,
\end{equation}
where $\ell$ is the ecliptic longitude, and $r$ the Earth-Sun radius with the semi-major axis length $a = 1 {\rm ~AU}$, the eccentricity $e$ and what is called the true anomaly $\nu$. The velocity, as an expansion to first order in $e$ is 
\begin{equation}
    \mathbf{V}_{\oplus}(t)= - V_\oplus (\sin L+e \sin (2 L-\varpi)) \hat{\boldsymbol{\epsilon}}_{\mathbf{x}} + V_\oplus (\cos L+e \cos (2 L-\varpi)) \hat{\boldsymbol{\epsilon}}_{\mathbf{y}} ,
\end{equation}
with the average Earth speed of $V_\oplus \approx 29.79 \mathrm{~km} / \mathrm{s}$, and using the same notation for the specific orbital parameters $L$, $g$, and $\varpi$ as in \cite{McCabe:2013kea}. 

The logic of our methodology is then as follows. From our core simulations, we obtain the \CNB phase-space for today, i.e. around $z \approx 0$. Due to the nature of discreteness of our simulations, this phase-space applies to a cubical region of space that contains the Earth-Sun system. We now apply the same procedure of our core \CNB simulations to obtain the phase-space for each day of the year: we initialize neutrinos at Earth's location for each day separately, run a simulation containing only the Earth-Sun system backwards one year, then use the phase-space we have from the core simulations as the end-point in Liouville's theorem, i.e.
\begin{equation}
    f_{\rm day}(\mathbf{x}_\oplus(z_{\rm day}), \mathbf{q}_j(z_{\rm day}), z_{\rm day}) = f(\mathbf{x}_\oplus, \mathbf{q}_j(0), 0), 
\end{equation}
where $z_{\rm day}$ is the redshift of the particular start day of the year. To connect optimally to the phase-space obtained from our core simulations, we conduct the daily simulations in an Earth-centered, MW-orientated frame, where the Sun is then moving with velocity $- \mathbf{v}_{\rm MW}$. The setup is illustrated in Fig. \ref{fig:mod_sketch}.

The extended cosmic durations of our core simulations, spanning from redshift $z = 4$ to the present day, required careful treatment of all DM gravitational forces. Over these longer time-scales, even the relatively weak gravitational effects of the MW DM halo accumulate to produce significant alterations in neutrino trajectories and phase-space distributions. However, the Earth-Sun simulations operate over much shorter time-scales of approximately one year, allowing us to neglect the weaker DM gravitational forces since their cumulative influence becomes negligible compared to the dominant solar gravitational effects. This can be quantified by comparing the gravitational accelerations: at Earth's orbital radius, the Sun produces $a_\odot \approx 6 \, \mathrm{mm/s^2}$, while the MW DM halo (assuming an NFW profile) contributes only $a_{\mathrm{MW}} \approx 10^{-7} \, \mathrm{mm/s^2}$. When integrated over the one-year duration of our Earth-Sun simulations, this weak DM acceleration has negligible impact on neutrino trajectories. Instead, the DM gravitational effects enter our framework through the phase-space distributions computed in the core simulations, which are then applied via Liouville's theorem as discussed above. This approach allows us to determine the local number density for each day throughout the year for the entire \CNB population, without the need to artificially separate bound and unbound neutrino components. The resulting number densities are technically computed in the MW rest frame, but since Earth's orbital velocities represent only small variations around $\mathbf{v}_{\mathrm{MW}} \ll c$, the corresponding Lorentz factors remain close to unity throughout the year.

In summary, the two-stage simulation approach allows us to properly account for effects at different scales: first capturing large-scale gravitational clustering effects over cosmological timescales, then modeling solar gravitational focusing over Earth's orbital period.

\section{Results}
\label{sec:results}

\begin{figure}[t!]
    \centering
    \begin{subfigure}{.488\textwidth}
        \centering
        \includegraphics[width=\linewidth]{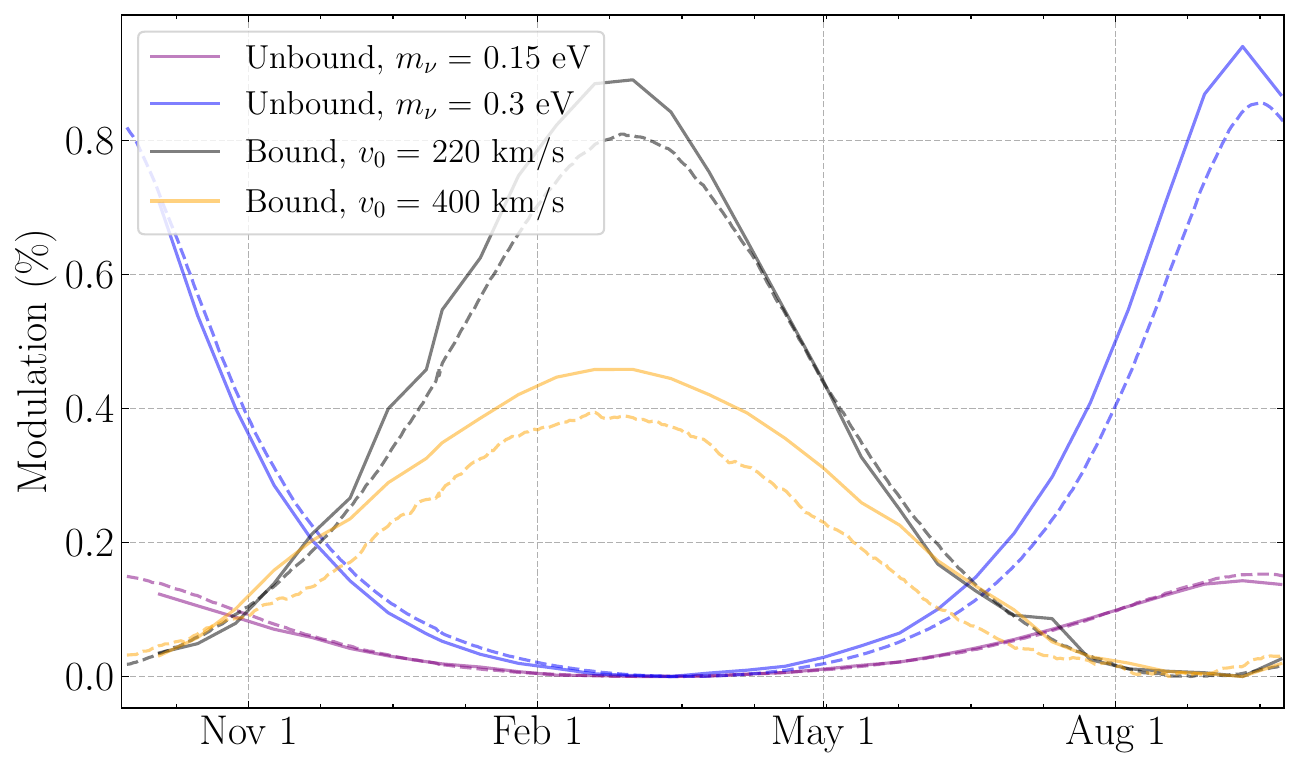}
    \end{subfigure}%
    \begin{subfigure}{.472\textwidth}
        \centering
        \includegraphics[width=\linewidth]{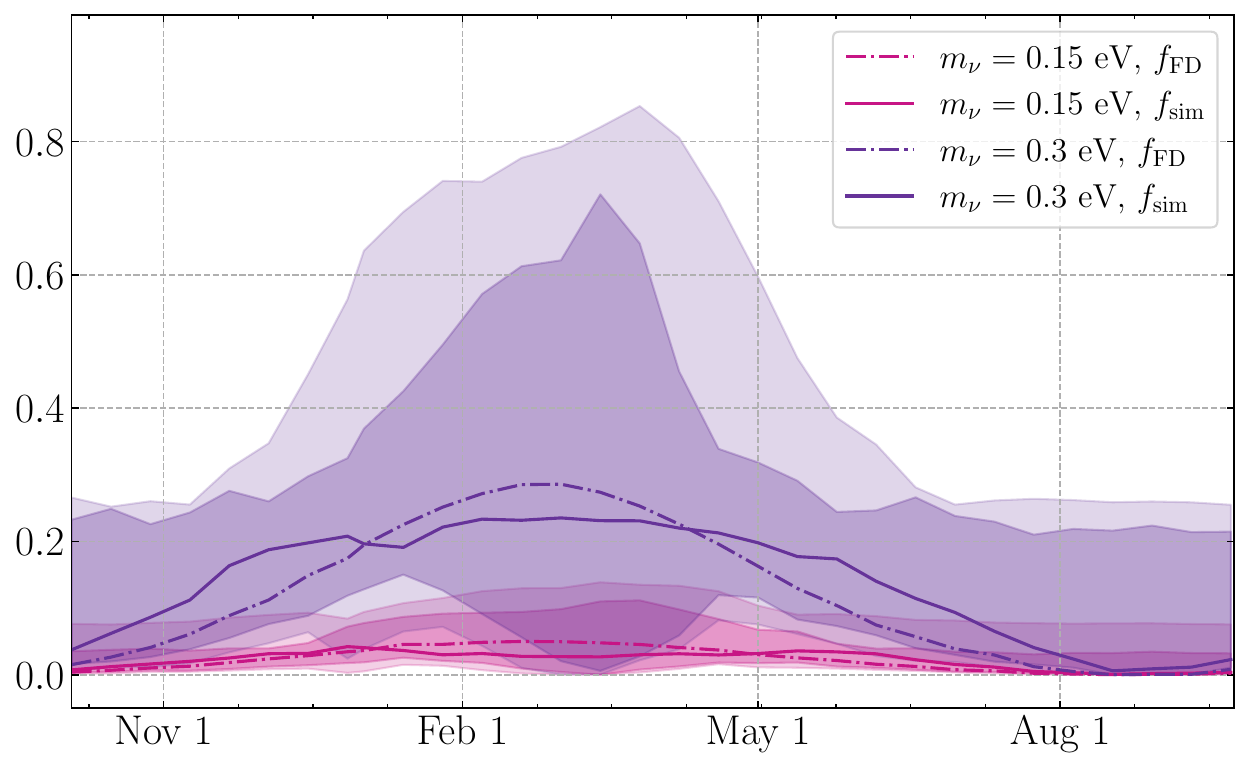}
    \end{subfigure}
    \caption{Annual modulation of the \CNB number density on Earth. Left panel: Comparison of numerical results (solid lines) with analytical predictions from \cite{Safdi:2014rza} (dashed lines) for both unbound and bound \CNB scenarios with neutrino masses of 0.15 eV and 0.3 eV. Right panel: Modulation rates from our full numerical treatment. Dark and light shaded regions show $68\%$ and $95\%$ confidence intervals across 26 dark matter halos. 
    Dash-dotted curves represent results using the distribution of the unbound case ($f_{\rm FD}$) with the relative velocity of the bound case ($\mathbf{v}_{\rm MW}$).
    For $0.3$ eV neutrinos, modulation varies from ${\sim}0\%$ to ${\sim}0.8\%$, while for $0.15$ eV neutrinos, it remains below ${\sim}0.1\%$.}
    \label{fig:mod_curves}
\end{figure}

In this section, we present the outcomes of our numerical simulations of \CNB annual modulation. We first validate our methodology against established analytical frameworks and then examine the modulation patterns predicted by our full numerical treatment. Specifically, we consider the impact of different DM halo morphologies on our results. Finally, we investigate which velocity ranges of the \CNB distribution contribute most significantly to the observed annual variations.

\paragraph{\CNB annual modulations.} Figure \ref{fig:mod_curves} presents our main findings. In the left panel, we validate our simulation methodology by comparing our numerically derived modulation curves with those from the analytical framework described in Section \ref{sec:ana_methods}. To reproduce the analytical scenarios, we simulate the Sun moving through our computational domain with a relative velocity of $\mathbf{v}_{\mathrm{C\nu B}}$ for the unbound case and $\mathbf{v}_{\rm MW}$ for the bound case, using the distributions $f_{\rm FD}$ and $f_{\rm SHM}$, respectively, as boundary conditions. The right panel shows results from our complete numerical treatment, where we use the phase-space from our core simulations (labeled with $f_{\rm sim}$) with $\mathbf{v}_{\rm MW}$ as the relevant relative velocity. Notably, our modulation curves peak around March 1, similar to the analytically predicted bound case, showing that the location of maximum amplitude is primarily determined by the relative velocity.

Our comprehensive numerical approach — incorporating non-linear gravitational clustering effects and treating the \CNB as a unified population rather than separate bound and unbound components — reveals significant variations in modulation amplitude in the relative sense. For $0.3$ eV neutrinos (the heaviest mass in our simulation framework), modulation peaks around March 12, reaching up to ${\sim}0.8\%$. However, this amplitude can be reduced to virtually $0\%$ depending on the specific DM halo morphology. For $0.15$ eV neutrinos, while the modulation maximum occurs around the same date, the overall rates remain modest, not exceeding ${\sim}0.1\%$ regardless of DM halo configuration.

In one of our earlier works we showed the fully bound percentage of the \CNB to be at most ${\sim}17\%$ for the highest $0.3$ eV neutrino mass, and this value drops below ${\sim}5\%$ for masses $\lesssim 0.15$ eV \cite{Zimmer:2023jbb}. 
Within the analytical framework, the combination of the distribution of the unbound case, $f_{\rm FD}$, with the relative velocity of the bound case, $\mathbf{v}_{\rm MW}$, should then give results closer to our predictions. There is indeed agreement with the median curves of our DM halo results when using this combination in our numerical setup.
We show this case with the dash-dotted modulation curves in the right panel of figure \ref{fig:mod_curves}.

\begin{figure}[t!]
    \centering
    \begin{subfigure}{.5\textwidth}
        \centering
        \includegraphics[width=\linewidth]{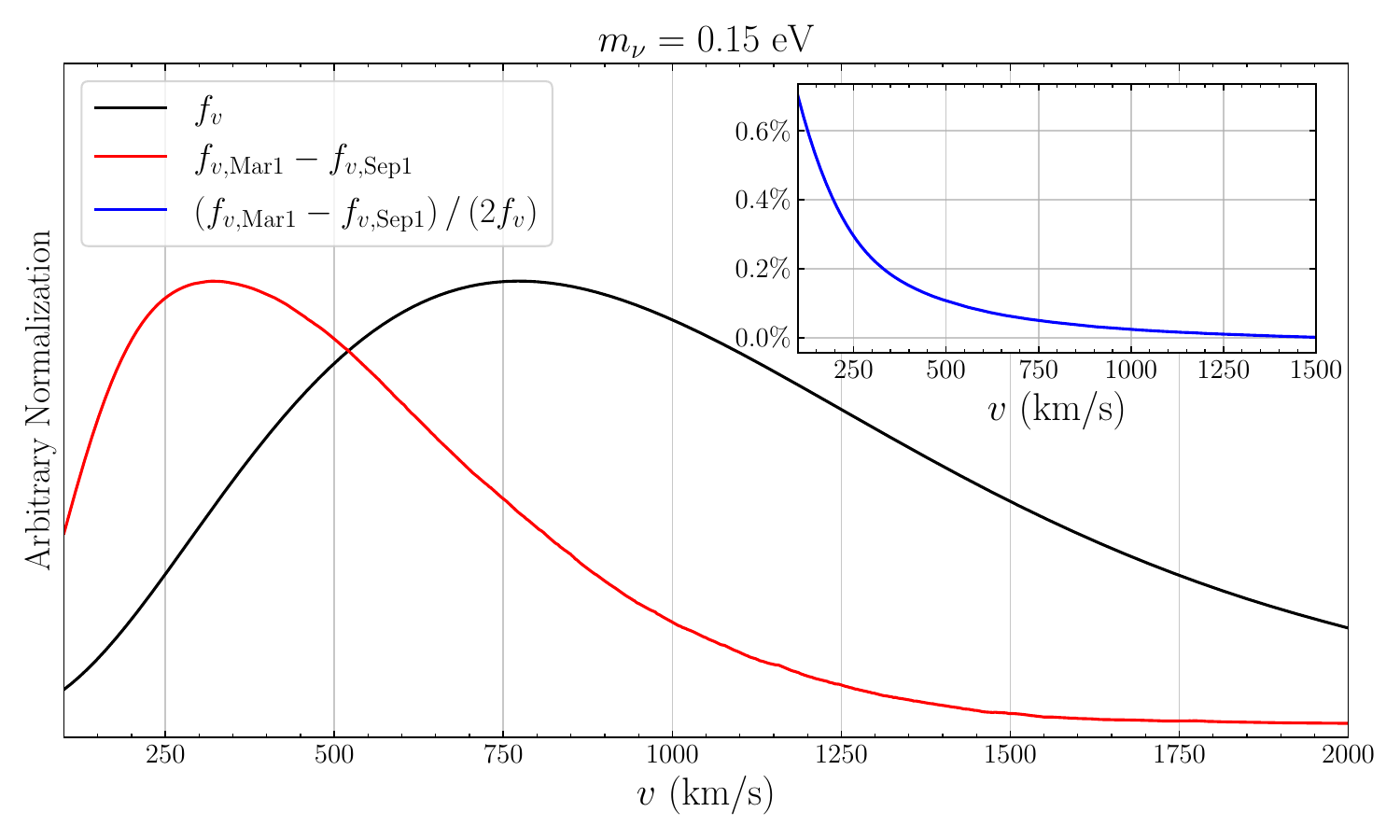}
    \end{subfigure}%
    \begin{subfigure}{.5\textwidth}
        \centering
        \includegraphics[width=\linewidth]{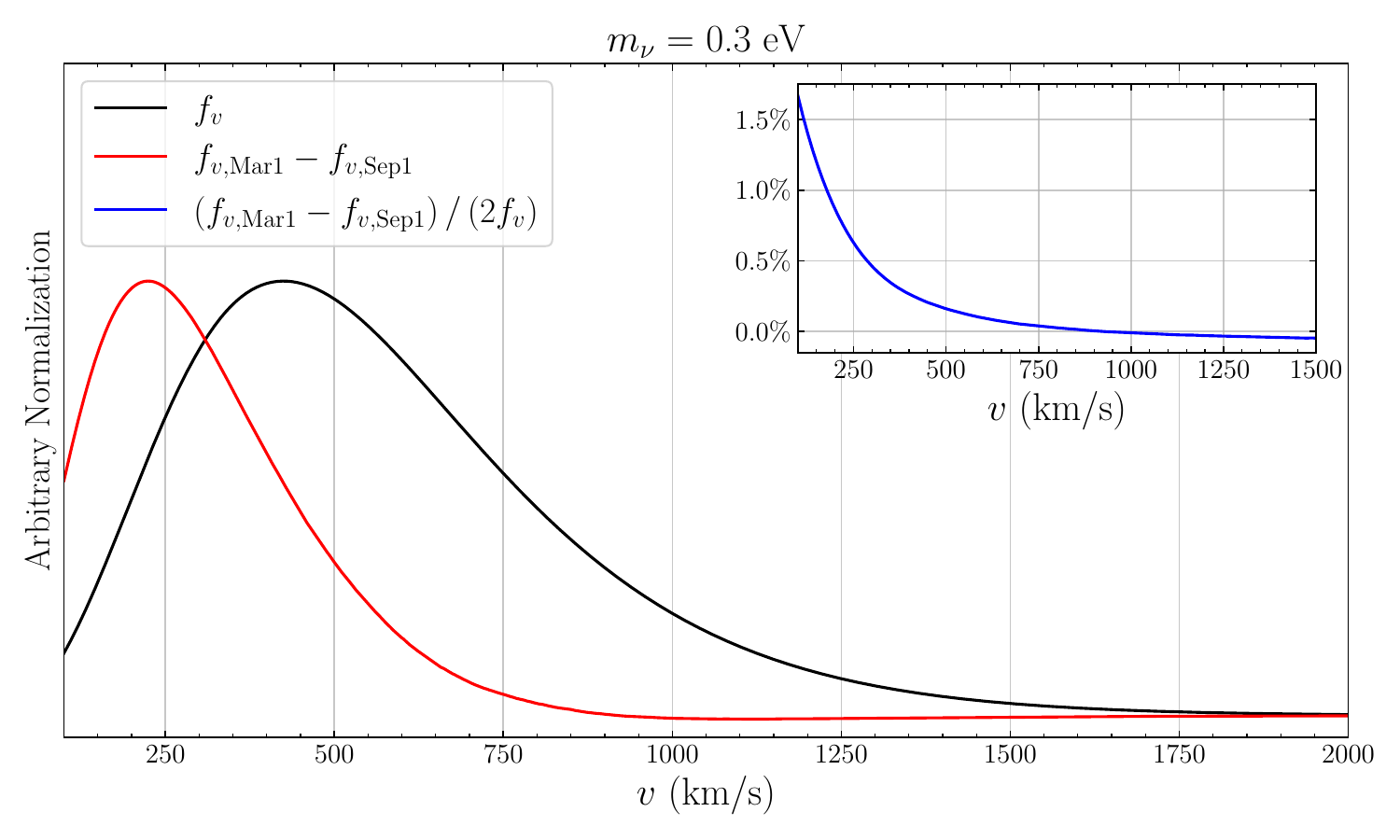}
    \end{subfigure}
    \caption{Velocity-dependent contributions to \CNB annual modulation for neutrino masses of $0.15$ eV (left panel) and $0.3$ eV (right panel). The black curves show the \CNB velocity distribution, and red curves the difference between March 1 and September 1 phase-spaces. Both curves are arbitrarily normalised to compare their shape. The inset blue curves show the percent modulation as a function of velocity. The mismatch between velocities most affected by modulation (red peak) and the most abundant velocities in the distribution (black peak) explains the limited overall modulation amplitude. All curves have been smoothed with a Savitzky-Golay filter to reduce some residual numerical noise.}
    \label{fig:vel_curves}
\end{figure}

\paragraph{\CNB velocity analysis.} To understand which \CNB velocities contribute most to annual modulation, we extend the velocity-dependent analysis approach of \cite{Lee:2013wza} from DM to \CNB detection. Within the relevant velocity window of $100$--$2000 \, {\rm km/s}$ for our benchmark masses, we define:
\begin{equation}
f_{v, t} \equiv \int d\Omega \, v^2 f(\mathbf{v}, t) \, .
\end{equation}
This integrates the phase-space density over solid angle for all velocity vectors of equal magnitude, providing insight into how different velocities contribute to modulation. Figure \ref{fig:vel_curves} displays this quantity for our benchmark masses ($0.15$ eV, left panel; $0.3$ eV, right panel) with normalized curves for shape comparison.
The black curves show the overall velocity dependence of the \CNB phase-space ($f_v$) for any one day, revealing that velocities around ${\sim}750 \, {\rm km/s}$ ($0.15$ eV) and ${\sim}400 \, {\rm km/s}$ ($0.3$ eV) dominate the phase-space density and thus the number density. The blue inset curves confirm that gravitational focusing most strongly affects slower neutrinos, showing the velocity-dependent modulation rate $(f_{v,{\rm Mar1}} - f_{v,{\rm Sep1}})/(f_{v,{\rm Mar1}} + f_{v,{\rm Sep1}})$ in percent. As the denominator varies little, it is shown as $2 f_v$ in figure \ref{fig:vel_curves} for convenience. The red curves highlight which neutrino velocities exhibit the greatest absolute difference between maximum (March 1) and minimum (September 1) modulation periods, proportional to $(f_{v,{\rm Mar1}} - f_{v,{\rm Sep1}})$. These peaks occur at lower velocities — around ${\sim}300 \, {\rm km/s}$ ($0.15$ eV) and ${\sim}200 \, {\rm km/s}$ ($0.3$ eV) — than the peaks of the overall \CNB velocity distribution.

\section{Discussion \& conclusions}
\label{sec:discussion_and_conclusions}

We have employed numerical simulations to investigate the annual modulation of the \CNB number density on Earth caused by the Sun's gravitational potential, comparing our results with previous analytical approaches. Our simulation setup is depicted in fig.~\ref{fig:mod_sketch}. This work, alongside our previous studies \cite{Zimmer:2023jbb,Zimmer:2024max}, demonstrates the versatility and efficiency of the N-1-body method for examining C$\nu$B-related questions across various gravitational scales. 

Our results are demonstrated by using two neutrino masses of $0.15$ eV and $0.3$ eV. Although both are compatible with laboratory constraints \cite{KATRIN:2021uub}, they are above the tight constraints obtained from cosmological data (see e.g. the newest data set of \cite{DESI:2024mwx}). We chose to proceed with these two masses for the following reasons. First, we wanted to have as close a comparison as possible to the previous analytical work of \cite{Safdi:2014rza} and their chosen heavy mass of $0.35$ eV. The heaviest mass our simulation framework is programmed to accommodate is $0.3$ eV. Second, there seem to be avenues to address the current tensions and anomalies in cosmology and the $\Lambda$CDM paradigm, which involve neutrinos that are heavier than the current model-dependent mass bounds (see e.g. \cite{Chacko:2019nej,FrancoAbellan:2021hdb,Escudero:2019gfk,Escudero:2020ped}). The last reason concerns the nature of our results: there is no significant annual modulation for neutrino masses $\lesssim 0.15$ eV, and there is nothing for us to demonstrate in the mass regime compatible with cosmological bounds. With this in mind, our key findings include:

\begin{itemize}
    \item Our methodology successfully reproduces modulation curves similar to those derived from the analytical framework in \cite{Safdi:2014rza} when artificially treating the \CNB as either fully unbound or fully bound to the MW (see left panel of fig. \ref{fig:mod_curves}).
    \item The modulation curve shape, particularly the location of maximum amplitude, is primarily determined by the relative velocity under consideration. In our case, the appropriate reference is the \CNB velocity relative to the MW, which enables proper connection to our phase-space values computed in the MW rest frame. Consequently, our modulation curve shape (right panel of fig. \ref{fig:mod_curves}) resembles the bound case predicted analytically.
    \item For our heaviest neutrino mass of $0.3$ eV, when incorporating the C$\nu$B's prior gravitational clustering onto the MW DM halo, the annual modulation amplitude varies considerably. Specifically, around March 12, the time of maximal modulation predicted analytically, the amplitude can vanish completely for some DM halo morphologies.
    This parallels our finding in \cite{Zimmer:2023jbb}, where numerical treatment of \CNB gravitational clustering due to DM halos revealed potential underdensities where analytical predictions indicated only overdensities. For lighter neutrinos ($\lesssim 0.15$ eV), modulation rates are negligible, reaching only ${\sim}0.1\%$ maximum. Nevertheless, these rates can still decrease around analytically predicted maxima due to DM halo asymmetry.
    \item The variation in modulation rates across different DM halos potentially connects to our findings in \cite{Zimmer:2023jbb}, where we showed that asymmetric DM halos can diffuse lower-momentum neutrinos, preventing them from reaching Earth — an effect not captured by previous analytical frameworks. As shown in fig. \ref{fig:vel_curves} (inset), the Sun's gravitational focusing is most effective on these slower neutrinos, explaining how different DM halo configurations affect modulation rates.
    \item Analysis of individual \CNB velocity bins (fig. \ref{fig:vel_curves}) clarifies why larger modulation rates than those found in this work are unlikely. For $0.3$ eV neutrinos, a $1\%$ modulation results from the Sun's gravitational focusing on ${\sim}200 \, {\rm km/s}$ velocities (blue curve, inset plot). However, these velocities occupy the lower tail of the \CNB phase-space distribution (black curve), despite today's non-relativistic nature of the C$\nu$B. This is more clearly shown by the fact that the difference in phase-spaces between time of maximal (March 1) and minimal (September 1) modulation (red curve) is highest for velocities in the lower tail of the \CNB velocity distribution. Essentially, the combination of peak \CNB velocities for a given neutrino mass, the Sun's mass, and Earth's distance and velocity relative to the Sun does not create optimal conditions for substantial annual modulation.
\end{itemize}

\noindent Our numerical investigation of \CNB annual modulation extends beyond previous analytical frameworks by incorporating parts of the gravitational history of relic neutrinos. We demonstrate that the modulation pattern is primarily shaped by the relative velocity between the \CNB and the MW, while its amplitude is influenced by DM halo morphology. The interplay between solar gravitational focusing (which predominantly affects slower neutrinos) and the actual velocity distribution of the \CNB (which peaks at higher velocities) fundamentally limits the overall modulation amplitude. This work highlights how numerical simulations across multiple gravitational scales can reveal subtle effects missed by idealized analytical approaches, particularly the potential suppression of modulation signals due to DM halo asymmetries.

\acknowledgments

The simulations for this work were performed on the Snellius Computing Clusters at SURFsara. This publication is part of the project ``One second after the Big Bang'' NWA.1292.19.231 which is financed by the Dutch Research Council (NWO). SA was partly supported by MEXT KAKENHI Grant Numbers, JP20H05850, JP20H05861, and JP24K07039.

\appendix


\providecommand{\href}[2]{#2}\begingroup\raggedright\endgroup

\end{document}